\documentclass[prd,amsmath,amssymb,preprint]{revtex4}
\usepackage{graphicx}% Iclude figure files                                    
\usepackage[english]{babel}
\usepackage{feynmf}
\usepackage{dcolumn}% Align table columns on decimal point                      
\usepackage{bm}% bold math                                                      
\usepackage{longtable}
\usepackage{pdflscape}
\usepackage{hyperref}
\usepackage{amsmath}
\renewcommand{\vec}[1]{{\mbox{\boldmath$#1$}}}
\begin{document}
%%%%%%%%%%%%%%%%%%%%%%%%%%%%%%%%%%%%%%%%%%%%%%%%%%%%%%%%%%%%%%%%%%%%%%%%
\title{Relativistic calculations of X-Ray transition energies and isotope
shifts in heavy atoms}
\author{I.~I.~Tupitsyn$^{1,2}$, N.~A.~Zubova$^{1,3}$, V.~M.~Shabaev$^{1}$, G.~Plunien$^{4}$, and Th. St\"ohlker$^{5,6,7}$}
\affiliation{$^1$Department of Physics, St. Petersburg State University,
7/9 Universitetskaya nab., St.~Petersburg 199034, Russia \\
$^2$ Center for Advanced Studies, Peter the Great St. Petersburg Polytechnic
University, Polytekhnicheskaja 29, St. Petersburg  195251, Russia\\
$^3$SSC RF ITEP of NRC ``Kurchatov Institute'', Bolshaya Cheremushkinskaya 25, Moscow, 117218, Russia\\
$^4$Institut f\"ur Theoretische Physik, TU Dresden, Mommsenstrasse 13, Dresden,
D-01062, Germany\\
$^5$GSI Helmholtzzentrum f\"ur Schwerionenforschung GmbH, D-64291 Darmstadt, Germany\\
$^6$Helmholtz-Institut Jena, D-07743 Jena, Germany\\
$^7$Institut f\"ur Optik und Quantenelektronik, Friedrich-Schiller-Universiat Jena, D-07743 Jena, Germany}
%%%%%%%%%%%%%%%%%%%%%%%%%%%%%%%%%%%%%%%%%%%%%%%%%%%%%%%%%%%%%%%%%%%%%%%%
\date{\today}
\begin{abstract}
X-Ray transition energies and isotope shifts in heavy atoms 
are evaluated. The energy levels
with vacancies in the inner shells are calculated within the approximation of the average of 
nonrelativistic configuration employing the Dirac-Fock-Sturm method. The obtained results are compared with other configuration-interaction theoretical calculations and with experimental data.
\end{abstract}
%%%%%%%%%%%%%%%%%%%%%%%%%%%%%%%%%%%%%%%%%%%%%%%%%%%%%%%%%%%%%%%%%%%%%%%%
\pacs{Valid PACS appear here}% PACS, the Physics and Astronomy
                             % Classification Scheme.
%Use showkeys class option if keyword
                              %display desired
\maketitle
\section{Introduction}
Precision calculations of energies of the X-ray emission lines and the related isotope 
shifts in heavy atomic systems are required by experiments \cite{Bearden_1967,
Deslattes_1985,Mooney_1992,Elliot_1996,Elliot_1998,Deslattes_2003}. The most
accurate to-date theoretical and experimental values of X-ray K-,L-,M- transition energies were tabulated in Ref. \cite{Deslattes_2003} and have been used in the NIST database \cite{NIST}. 
As to the isotope shift in heavy neutral atoms, first measurements of the isotope shifts
in X-ray K$\alpha_1$ lines for neutral uranium
isotopes have been performed by Brockmeier and co-authors
\cite{Brockmeier_1965} and for molybdenum isotopes 
by Sumbaev an Mezentsev \cite{Sumbaev_1965}. In Ref. \cite{Makarov_1996}, the experimental 
and theoretical study of the isotope shifts in X-ray L~lines
in neutral uranium was carried out. The isotope shifts of atomic
X-ray K~lines in mercury (Hg) were measured for different pairs of isotopes
in Ref. \cite{Lee_1978}.  

From the theoretical side, the binding energies in many-electron 
atoms can be calculated very accurately using the
multiconfiguration Dirac-Fock method (MCDF) \cite{Deslattes_2003,Indelicato_1998,Li_2012,Naze_2014} or configuration-interaction
Dirac-Fock-Sturm (CI-DFS) method \cite{Tupitsyn_2003, Tupitsyn_2005}.
But, as shown in Ref. \cite{Deslattes_2003}, the MCDF method is not efficient enough
for calculations of the inner-shell hole states. 
So, to take into account the correlation and Auger shift corrections to X-ray
lines, 
in Refs. \cite{Deslattes_2003,Indelicato_1998} the relativistic many-body
perturbation theory (RMBPT) was employed. We note also that in Ref. \cite{Deslattes_2003} the quantum electrodynamics (QED) corrections have been determined using Welton's approximation.  

In the present paper we use the assumption that the center of gravity of the
X-ray emission line in heavy atoms can be calculated as the difference
of the averages of nonrelativistic valence configurations with
the different vacancies in the inner shells. This approximation
is used in the Dirac-Fock and CI-DFS calculations in this work. In this approach
the energy is averaged over all atomic terms of the nonrelativistic
valence configuration. The idea of the nonrelativistic configurational average
(``LS-average'') in the relativistic Dirac-Fock calculations was proposed in
\cite{Desclaux_1971, Lindgren_1974}. The validity of this approximation
is demonstrated by our calculations of the binding energies of X-ray lines.

To calculate the Auger shifts we use the RMBPT method but, in contrast to Ref.
\cite{Deslattes_2003}, in the Brillouin-Wigner form. The obtained non-QED
results are combined with the corresponding QED contributions, which have been
evaluated by including the model Lamb-shift operator into the Dirac-Coulomb-Breit Hamiltonian
\cite{Shabaev_2013,Shabaev_2015,Tupitsyn_2016}. As the result, the most precise 
theoretical predictions for the energies and isotope shifts of X-ray K and
L~lines are  presented.  

The atomic units ($\hbar=m=e=1$) are used throughout the paper. 
%%%%%%%%%%%%%%%%%%%%%%%%%%%%%%%%%%%%%%%%%%%%%%%%%%%%%%%%%%%%%%%%
\section{Method of calculation}

In order to calculate the X-ray transition energies we use the following
three-step large-scale CI-DFS method \cite{Tupitsyn_2003,Tupitsyn_2005}.
At the first step, to obtain the one-electron wave functions for
the occupied atomic shells, we use the Dirac-Fock method \cite{Bratsev_1977} with
the average of nonrelativistic configuration. Then the DFS orbitals are obtained
by solving the DFS equations for the vacant shells. At the last step, the relativistic CI+MBPT
method is used to obtain the many-electron wave functions and the total energies.
%
%%%%%%%%%%%%%%%%%%%%%%%%%%%%%%%%%%%%%%%%%%%%%%%%%%%%%%%%%%%%%%%%
\subsubsection*{Average of nonrelativistic configuration.
``LS-average"}
%%%%%%%%%%%%%%%%%%%%%%%%%%%%%%%%%%%%%%%%%%%%%%%%%%%%%%%%%%%%%%%%
To evaluate the transition energies with vacancies in the inner shells
we use the CI-DFS method in the approximation of the average of
nonrelativistic configuration (for more details, see, the Ref.
\cite{thesis_Tupitsyn}). The choice of this approach for the case of an
atom with open shells is caused by the following reason. The expression
for the energy in one-configuration Dirac-Fock method for atoms with open
nonrelativistic shells does not converge to the corresponding non-relativistic
expression if the speed of light tends to infinity. In other words,
the one-configuration Dirac-Fock method corresponds to the $jj$-scheme of coupling,
which in its pure form is almost never realized in neutral atoms with open
valence shells, and does not lead to the $LS$-coupling scheme (Russell-Saunders coupling) in
the nonrelativistic limit. To remedy this shortcoming, it is necessary to
consider the interaction  of the relativistic configurations that correspond
to the same nonrelativistic one. This corresponds to the intermediate type of
coupling or the approximation of the barycenter of the nonrelativistic
configuration.

The X-ray emission line widths of heavy atoms are so large that they can
exceed the value of the multiplet splitting of the atomic valence levels.
In this case, to calculate the position of the center of gravity (or maximum)
of the X-ray line observed in the experiment, it is sufficient to calculate
the transition energies and isotope shifts in the nonrelativistic
configuration average approximation.

The idea of the configuration average in the case of nonrelativistic
Hartree-Fock method  was treated in detail by Slater
\cite{Slater_1960}. The formalism can easily be extended to include also 
the average of several relativistic configurations \cite{Lindgren_1974}
corresponding to the same nonrelativistic one in the Dirac-Fock
calculations.  This configurational averaging technique was named as 
nonrelativistic ``LS-average".

%%%%%%%%%%%%%%%%%%%%%%%%%%%%%%%%%%%%%%%%
Let the nonrelativistic shells are enumerated by indices $A$ and $B$ which incorporate
the quantum numbers $n_a l_a$  and $n_b l_b$, respectively, and the relativistic shells are numbered
by indices $a$ and $b$. In the approximation of the barycenter of
nonrelativistic configuration the expression for the Dirac-Fock 
energy is given by
%%%%%%%%%%%%%%%%%%%%%%%%%%%%%%%%%%%%%%%%%
%
\begin{equation}
E^{\mathrm{DF}}_{\mathrm{nav}}=\sum_{a} \tilde q_a I_a+
\frac{1}{2} \sum_{a,b} W_{A,B} \, \sum_{\mu_a=-j_a}^{j_a} 
\sum_{\mu_b=-j_b}^{j_b}
[\langle a \mu_a,b\mu_b|a\mu_a,b\mu_b \rangle -
\langle a \mu_a,b\mu_b|b\mu_b, a\mu_a \rangle],
\label{E_nav1}
\end{equation}
where $q_{A}$ is the number of electrons (occupation number) in the
nonrelativistic shell $A$, $\widetilde q_a$ is the
average occupation number of the relativistic subshell $a$,
\begin{equation}
\widetilde q_a=\frac{2j_a+1}{4l_A+2} \, q_{A} \,,
\label{occ1}
\end{equation}
$I_a$ is the one-electron diagonal matrix element of the Dirac operator
$\hat h_{\rm D}$, which is independent of the projection $\mu$,
\begin{equation}
I_a=\langle a \mu  | \hat{h}_{\rm D} |a \mu \rangle,
\end{equation}
and
\begin{equation}
W_{A,B}=
\begin{cases} \displaystyle
\frac{q_{A} \, q_{B}}{(4l_A+2)(4l_B+2)} \,, \qquad A \neq B 
\\ \displaystyle
\frac{q_{A} \, (q_{A}-1)}{(4l_A+2)(4l_A+1)}\,, \qquad A = B. 
\end{cases}
\end{equation}
The detailed formulas for the Dirac-Fock energy in the approximation of the
average of nonrelativistic configuration are given in Appendix.
%
%%%%%%%%%%%%%%%%%%%%%%%%%%%%%%%%%%%%%%%%%%%%%%%%%%%%%%%%%%%%%%%%
\subsubsection*{CI-DFS method with average of nonrelativistic configuration}
%%%%%%%%%%%%%%%%%%%%%%%%%%%%%%%%%%%%%%%%%%%%%%%%%%%%%%%%%%%%%%%%%%%%%%%%%
%
To take into account the electron correlations the large-scale 
configuration-interaction (CI) method in the basis of four-component Dirac-Fock-Sturm (DFS)
orbitals $\varphi_a$ is used. These orbitals are obtained by solving the
Dirac-Fock-Sturm equations \cite{Tupitsyn_2003,Tupitsyn_2005}. Various
excited configurations are obtained from the main configuration by single and 
double excitations of ``active'' electrons. According to the method 
of group functions \cite{McWeeny_1992}, the wave functions are presented in the form of an
antisymmetric product of the wave functions of two groups of electrons.
The first one is the group of ``active''
electrons, while the second one is the group of ``frozen'' electrons.
In the formulation of our problem the ``active'' are the core electrons,
and the ``frozen''  are the valence electrons. The interaction with the valence
electrons is taken into account by the introduction of a single-particle
potential, which is the sum of the Coulomb and exchange potentials. The Coulomb
and exchange potentials of the valence electrons are constructed in the standard
way using the first order reduced density matrix taken in the approximation
of the average of nonrelativistic valence configuration,
\begin{equation}
\label{dens_matrix}
\rho^{(\rm val)}(\vec{r},\vec{r}^\prime)=\mathop{{\sum_a}^{(\rm val)}} 
\frac{\widetilde q_a }{2j_a+1} \, \sum_{\mu=-j_a}^{j_a} 
\varphi_{a \mu}(\vec{r}) \varphi^{+}_{a \mu}(\vec{r}^{\prime}) \,,
\end{equation}
where the summation runs over indices of the valence electrons and
$\widetilde q_a$ is defined by Eq. (\ref{occ1}).
%
%%%%%%%%%%%%%%%%%%%%%%%%%%%%%%%%%%%%%%%%%%%%%%%%%%%%%%%%%%%%%%%%
\subsubsection*{QED corrections}
%%%%%%%%%%%%%%%%%%%%%%%%%%%%%%%%%%%%%%%%%%%%%%%%%%%%%%%%%%%%%%%%%%%%%%%%%
In this paper we approximate the QED potential by the 
following sum
\begin{equation}
V^{\rm QED} = V^{\rm SE} + V^{\rm Uehl} +  V^{\rm WK} \,,
\end{equation}
where $V^{\rm SE}$ is so-called model self-energy operator, 
$V^{\rm Uehl}$ and $V^{\rm WK}$ are the Uehling and Wichmann-Kroll
parts of the vacuum polarization, respectively.
Both $V^{\rm Uehl}$ and $V^{\rm WK}$ are local potentials.
The Uehling potential can be evaluated by a direct numerical integration
of the well-known formula \cite{Uehling_1935} or,
more easily, by using the approximate formulas from Ref.
\cite{Fullerton_1976}. A direct numerical evaluation of the Wichmann-Kroll
potential $V^{\rm WK}$ is rather complicated. For the purpose of the
present work, it is sufficient to use the approximate formulas for this potential 
from Ref.~\cite{Fainshtein_1991}.

Following Refs. \cite{Shabaev_2013, Shabaev_2015} we represent the one-electron SE
operator as the sum of local $V^{\rm SE}_{\rm loc}$ and nonlocal 
$V_{\rm nl}$ parts,
\begin{eqnarray}
V^{\rm SE} = V^{\rm SE}_{\rm loc} + V_{\rm nl} \,,
\label{se1}
\end{eqnarray}
where the nonlocal potential is given in a separable form,
\begin{eqnarray}
V_{\rm nl} =
\sum_{i,k=1}^{n} |\phi_i\rangle B_{ik}\langle \phi_k|.
\label{nl1}
\end{eqnarray}
Here $\phi_i$ are so-called projector functions. The choice of these
functions is described in details in Ref. \cite{Shabaev_2013}.
%for the present case.
%The matrix elements $B_{ik}$ satisfy the following property:
The constants $B_{ik}$ are chosen so that the matrix elements of the
model operator $V^{\rm SE}_{ik}$ calculated with hydrogenlike wave
functions $\psi_i$ are
%exactly coincide with
equal to the matrix elements $Q_{ik}$ of the exact
SE operator $\Sigma(\varepsilon)$
\cite{Shabaev_1993}:
\begin{eqnarray}
\langle \psi_i | V^{\rm SE} | \psi_k \rangle = Q_{ik} \equiv \frac{1}{2}\langle \psi_i | \, \left
[ \Sigma(\varepsilon_i)+ \Sigma(\varepsilon_k) \right ]\,| \psi_k \rangle .
\end{eqnarray}
Introducing two matrices,
$\Delta Q_{ik}=Q_{ik}- \langle \psi_i|V^{\rm SE}_{\rm loc}| \psi_k\rangle$
and $D_{ik} = \langle \phi_i|\psi_k\rangle $,
we find that
\begin{eqnarray}
B_{ik} = \sum_{j,l=1}^{n} (D^{-1})_{ji}
\langle \psi_j| \Delta Q_{jl} | \psi_l\rangle (D^{-1})_{lk} \,.
\label{nl2}
\end{eqnarray}
The local part of the SE potential was taken in
a simple form \cite{Shabaev_2013},
\begin{eqnarray}
 V^{\rm SE}_{{\rm loc},\kappa}(r) =  A_{\kappa} \exp{(-r/\lambdabar_C)}\,,
\label{local}
\end{eqnarray}
where the constant $A_{\kappa}$ is chosen to reproduce
the  SE shift for the lowest energy level at the given $\kappa$
in the corresponding H-like ion and $\lambdabar_C=\hbar/(mc)$.
The computation code based on this method is presented in
Ref. \cite{Shabaev_2015}.
%%%%%%%%%%%%%%%%%%%%%%%%%%%%%%%%%%%%%%%%%%%%%%%%%%%%%%%%%%%%%%%%%%%%%%%%%
\section{Energies of X-ray emission lines}
%%%%%%%%%%%%%%%%%%%%%%%%%%%%%%%%%%%%%%%%%%%%%%%%%%%%%%%%%%%%%%%%%%%%%%%%%
In Table \ref{G_U}, the natural widths taken from Ref. \cite{Campbell_2001}
are compared with the widths of the multiplet splitting for X-ray lines
in uranium. The multiplet splitting arises if the
atom contains open valence shells.
When a core electron vacancy is created, an unpaired electron in the core can couple 
with electrons
the in outer shells. This creates a number of states which can be
seen in photoelectron spectrum as a multi-peak envelope.

The comparison of the widths gives an indication of the right application
of the approximation of the barycenter of nonrelativistic configuration.
It is expected that the approximation of the barycenter configuration
is applicable in the case when the natural linewidth is bigger than or
at least comparable to the multiplet splitting magnitude.
The data in Table~\ref{G_U} demonstrate that the required conditions are 
fulfill.
\begingroup
%\squeezetable
%%%%%%%%%%%%%%%%%%%%%%%%%%%%%%%%%%%%%%%%%%%%%%%%%%%%%%%%%%%%%%%%%%%%%%%%%
\begin{table}[ht]
\caption{\small The comparison of the natural line widths and 
the widths of the multiplet splitting for uranium X-ray lines.
$\Delta$M is the width of the multiplet splitting and $\Gamma$ is the natural line width.}
\begin{center}
\vspace{-2.0mm}
\begin{tabular}{c|c|c|c}
\hline
&&&\\[-3mm]
Line~ & Transition & ~~$\Gamma$ (eV)~~ & $\Delta$M (eV)  \\
\hline
&&&\\ [-6mm]
L$\alpha_2$  &~~2p$_{3/2}^{-1} \to $ 3d$_{3/2}^{-1}$~~ & 11.7  &  16.8 \\ [-1mm]
L$\beta_1$   & 2p$_{1/2}^{-1} \to $ 3d$_{3/2}^{-1}$
             & 13.5 & 16.55 \\ [-1.mm]
L$\beta_3$   & 2s$_{1/2}^{-1} \to $ 3p$_{3/2}^{-1}$ & 23.9  &  28.4  \\ [-1mm]
L$\beta_4$   & 2s$_{1/2}^{-1} \to $ 3p$_{1/2}^{-1}$ & 30.1  &  27.7  \\ [-1mm]
K$\alpha_1$  & 1s$_{1/2}^{-1} \to $ 2p$_{3/2}^{-1}$ & 104.5 &  27.7 \\  [-1mm]
K$\alpha_2$  & 1s$_{1/2}^{-1} \to $ 2p$_{1/2}^{-1}$ & 106.3 &  27.6 \\
\hline
\end{tabular}
\end{center}
\label{G_U}
\end{table}
%%%%%%%%%%%%%%%%%%%%%%%%%%%%%%%%%%%%%%%%%%%%%%%%%%%%%%%%%%%%%%%%%%%%%%%%%
The results of the calculations of the  K$\alpha$ lines for uranium,
xenon, and mercury
and the L~lines for uranium are presented in Tables \ref{EK}, \ref{Xe},
\ref{Hg}, and \ref{EL}, respectively. The calculations have been performed
using the Dirac-Fock method \cite{Bratsev_1977} in the approximation of the
barycenter of nonrelativistic configuration (\ref{E_nav1}) including 
the Breit, electron correlation, QED, and nuclear recoil (mass shift) contributions.
The nuclear charge distribution was taken into account within the
Fermi model with the root-mean-square nuclear radii taken from Ref. \cite{Kozhedub_2008,Angeli_2013}. 
The QED contributions are evaluated by including the model Lamb-shift operator into the
Dirac-Coulomb-Breit Hamiltonian \cite{Shabaev_2013}. 

The nuclear recoil effect is calculated within the Breit approximation using the relativistic nuclear recoil 
Hamiltonian \cite{Shabaev_1985,Palmer,Shabaev_1988,Shabaev_1998,Tupitsyn_2003},
\begin{eqnarray}   
\label{HM}
H_M &=& \frac{1}{2M}\sum_{i,k}\Bigl[
{\vec{p_i}}\cdot {\vec{p_k}} -\frac{\alpha Z}{r_i}\Bigl[\vec{\alpha_i}+
\frac{(\vec{\alpha_i}\cdot\vec{r_i})\vec{r_i}}{r_i^2}\Bigr]\cdot\vec{p_k}
\Bigr] \,.
\end{eqnarray} 
The uncertainties of the total values of the X-ray lines in Tables \ref{EK}, \ref{Xe},
\ref{Hg}, \ref{EL} are mainly due to the correlation and Auger shift
contributions which depend 
on the way of the calculations. The results of these calculations are 
unstable within 1 eV, so the conservative estimates of the uncertainty of
the order of 2-3 eV are used. In case of uranium atom, the nuclear polarization
and deformation corrections were taken from Refs. \cite{Plunien_1996,Nefiodov_1996,Volotka_2014} and \cite{Kozhedub_2008}, respectively. The uncertainty
of 50\% was assumed for these corrections. For $^{136}$Xe and $^{204}$Hg
atoms the nuclear polarization and deformation corrections 
are negligible \cite{Yerokhin_2015}.

The comparison of the energies of the K$\alpha$ lines for
$^{238}$U, $^{136}$Xe, and $^{204}$Hg and the L~lines for $^{238}$U with 
other theoretical results and
experimental data demonstrates very good agreement. 
This allows us to conclude that the approximation of the barycenter
of the nonrelativistic  configuration in the calculations of the X-ray
transition energies is applicable for heavy atoms with open
valence shells.
%%%%%%%%%%%%%%%%%%%%%%%%%%%%%%%%%%%%%%%%%%%%%%%%%%%%%%%%%%%%%%%%%%%%%%%%%%%
\begin{table}[ht]
\begin{center}
\small
\vspace{1mm}
\caption{\small Individual contributions to the energy of the K$\alpha$ lines
for $^{238}$U (in keV) with the nuclear charge radius R=5.8569 $\text{fm}$ in this work and R=5.8625 fm in Refs. \cite{Deslattes_2003,Indelicato_1992}.}
\vspace{2mm}
%\begin{ruledtabular}                    
\begin{tabular}{l|cc|cc}
\hline \hline
%\hspace{42mm}
\multicolumn{1}{c|}{ Transition} & \multicolumn{2}{c|}
{\hspace{3mm} K${\alpha_1}$ \hspace{3mm}} & 
\multicolumn{2}{c}{ \hspace{3mm} K${\alpha_2}$ \hspace{3mm}}
\\[-2mm]
&  This work & Theory \cite{Indelicato_1992} & This work & Theory \cite{Indelicato_1992}  \\[1mm]
\hline \hline
Dirac-Fock                 &  99.1031  &  99.1016~ & 95.2777~ &  95.2763~ \\[-1mm]
Breit                      &  -0.4339  &  -0.4319  & -0.3940  &  -0.3923 \\[-1mm]
Frequency-dependent Breit  &  ~0.0067  &  ~0.0066  & ~0.0126  &   0.0125 \\[-1mm]
QED                        &  -0.2466  &  -0.2436  & -0.2486  &  -0.2460 \\[-1mm]
Electron correlations + Auger shift   &   ~0.0038  & ~0.0030  & ~0.0030 & 0.0046 \\[-1mm]
Mass shift                 &  -0.0001  &   -       & -0.0001  &     -    \\[-1mm]
Nuclear polarization       &  0.0002  &   ~0.0002$^a$ &   0.0002      &  ~~0.0002$^a$   \\[-1mm]
Nuclear deformation        &  0.0001  &  -    &   0.0001        &  -        \\
Total                      & ~ 98.4333(38) ~ &  98.4359$^b$ & ~94.6508(30) & ~94.6553$^b$  \\[-1mm]
\hline
\multicolumn{1}{l|}{Theory \cite{Deslattes_2003}} & \multicolumn{2}{c|}{98.4336(36)}
 & \multicolumn{2}{c}{94.6531(37)}\\[-1mm]
\multicolumn{1}{l|}{Experiment \cite{Deslattes_2003,NIST}} & 
\multicolumn{2}{c|}{98.43158(28)} & \multicolumn{2}{c}{94.65084(56)}\\
\hline
\hline
\multicolumn{2}{l}{\small $^a$ Corrected according to Refs. \cite{Plunien_1996,Nefiodov_1996,Volotka_2014}.}\\[-3mm]
\multicolumn{5}{l}{\small $^b$ Corrected for the updated value of the nuclear polarization.}\\[-3mm]
\end{tabular}
\label{EK}                     
\end{center}
\end{table}
%%%%%%%%%%%%%%%%%%%%%%%%%%%%%%%%%%%%%%%%%%%%%%%%%%%%%%%%%%%%%%%%%%%%%%%%%%%
\begin{table}[ht]
\begin{center}
\small
\vspace{1mm}
\caption{\small Individual contributions to the energy of the
K$\alpha$-lines for $^{136}$Xe (in keV) with the nuclear charge radius equal
to R=4.7964 $\text{fm}$.}
\vspace{2mm}
%\begin{ruledtabular}                    
\begin{tabular}{c|cc|cc}
\hline \hline
%\hspace{42mm}
\multicolumn{1}{c|}{ Transition} & \multicolumn{2}{c|}
{\hspace{3mm} K${\alpha_1}$ \hspace{3mm}} & 
\multicolumn{2}{c}{ \hspace{3mm} K${\alpha_2}$ \hspace{3mm}}
\\[-2mm]
       &  This work & Theory \cite{Mooney_1992} & This work & Theory \cite{Mooney_1992}  \\[1mm]
\hline \hline
Dirac-Fock                 &  29.8909  &  29.8908~ & 29.5665~ &  29.5660~ \\[-1mm]
Breit                      &  -0.0736  &  -0.0733  & -0.0693  &  -0.0691 \\[-1mm]
Frequency-dependent Breit  &  ~0.0004  &  ~0.0004  &  0.0008  &   0.0008 \\[-1mm]
QED                        &  -0.0410  &  -0.0410  & -0.0416  &  -0.0416 \\[-1mm]
Electron correlations + Auger shift   &  ~0.0021   & ~0.0017  & ~0.0020 & 0.0022 \\[-1mm]
Mass shift & -0.0001 &- & -0.0001 & -\\[-1mm]
Total    & ~ 29.7788(21) ~ &  29.7787~~& ~29.4582(20) & ~29.4584  \\[-1mm]
\hline
\multicolumn{1}{c|}{Theory \cite{Deslattes_2003}} & \multicolumn{2}{c|}{29.7783(29)}
 & \multicolumn{2}{c}{29.4584(30)}\\
\multicolumn{1}{c|}{Experiment \cite{Deslattes_2003,NIST}} & 
\multicolumn{2}{c|}{29.77878(10)} & \multicolumn{2}{c}{29.458250(50)}\\
\hline
\hline
%\multicolumn{3}{l}{\small $^a$ Numbers in parentheses indicate unscreened (H-like) QED contributions.}\\[-3mm]
\end{tabular}
\label{Xe}                        
\end{center}
\end{table}
%%%%%%%%%%%%%%%%%%%%%%%%%%%%%%%%%%%%%%%%%%%%%%%%%%%%%%%%%%%%%%%%%%%%%
\begin{table}[ht]
\begin{center}
\small
\vspace{1mm}
\caption{\small Individual contributions to the energy of the 
K$\alpha$-lines for $^{204}$Hg (in keV) with the nuclear charge radius R=5.4744 $\text{fm}$.}
\vspace{2mm}
\begin{tabular}{c|c|c}
\hline \hline
\hspace{42mm}
& \hspace{9mm} K${\alpha_1}$ \hspace{9mm}
               & \hspace{9mm} K${\alpha_2}$ \hspace{9mm} \\
\hline && \\[-5mm]
Dirac-Fock                &   71.2322~  &   69.2850~    \\[-1mm]
Breit                     &   -0.2674   &  -0.2465      \\[-1mm]
Frequency-dependent Breit &   ~0.0034   &  ~0.0061      \\[-1mm]
QED$$                     &   -0.1519   &  -0.1540      \\[-1mm]    
Electron correlations + Auger shift  &   0.0029       &   0.0035        \\[-1mm]
Mass shift                &   -0.0001   &  -0.0001      \\[-1mm]    
{Theory} (this work)                 & 70.8191(18)    & 68.8942 (19)   \\[-1mm]
{Theory} \cite{Deslattes_2003}       & 70.8190(22)    & 68.8943 (23)   \\[-1mm]
{Experiment}  \cite{Deslattes_2003}  & 70.8195(18)    & 68.8951 (17) \\
\hline
\hline
\end{tabular}
\label{Hg}
\end{center}
\end{table}
%%%%%%%%%%%%%%%%%%%%%%%%%%%%%%%%%%%%%%%%%%%%%%%%%%%%%%%%%%%%%%%%%%%%%
\begin{table}[ht]
\begin{center}
\small
\vspace{1mm}
\caption{\small Individual contributions to the energy of the L-lines for
$^{238}$U (in keV) with the nuclear charge radius R=5.8569 $\text{fm}$.}
\vspace{2mm}
\begin{tabular}{c|c|c|c|c}
\hline \hline
\hspace{10mm}
              & \hspace{3mm} {\small L${\alpha_2}$} \hspace{3mm}
               & \hspace{3mm} {\small L${\beta_1}$} \hspace{3mm}
               & \hspace{3mm} {\small L${\beta_3}$} \hspace{3mm}
               & \hspace{3mm} {\small L${\beta_4}$} \hspace{3mm} \\[1mm]
\hline
Dirac-Fock                 &  13.4869  &  17.3123~ &   17.5446 &  16.6560 \\[-1mm]
Breit                      &  -0.0496  &  -0.0895  &  -0.0474  &  -0.0391 \\[-1mm]
Frequency-dependent Breit  &  ~0.0056  &  -0.0003  &  -0.0022  &  -0.0006 \\[-1mm]
QED                        &  -0.0058  &  -0.0037  &  -0.0401 &  -0.0404  \\[-1mm]
%     &    &  -0.0026   &  -0.0328   &  -0.0389  \\[-1mm]
%                  &  (-0.0090) &  (-0.0071)  &  (-0.0468) &  (-0.0473)   \\[-1mm]
Electron correlations + Auger shift &  0.0007 & 0.0010 & 0.0003 & 0.0002 \\[-1mm]
Mass shift                 &  -0.0000  &  -0.0000  &  -0.0000 &  -0.0000  \\[-1mm]
Theory  (this work) & 13.4379(17)  & 17.2198(20) & 17.4552(16) & 16.5762(16) \\[-1mm]
Theory \cite{Deslattes_2003} & 13.4382(14) & 17.2187(16) & 17.4565(36) & 16.5762(34)\\  
Experiment   \cite{Deslattes_2003,NIST} & 13.43897(19) & 17.22015(28) & 17.45517(73) & 16.57551(30)\\
\hline
\hline
%\multicolumn{5}{l}{\small $^a$ Nuclear polarization contribution }\\[-3mm]
\end{tabular}
\label{EL}
\end{center}
\end{table}
%%%%%%%%%%%%%%%%%%%%%%%%%%%%%%%%%%%%%%%%%%%%%%%%%%%%%%%%%%%%%%%%%%%%%%%%%%%%
\newpage
\section{Isotope shifts of X-ray lines in neutral uranium and mercury}
Isotope shifts of atomic systems give a useful tool for determination of the nuclear charge
radius differences (see, e.g., Refs. \cite{Elliot_1996,Schuch_2005,Brandau_2008,Orts_2006,Kozhedub_2008} and references therein).
For the last years a significant progress was gained in calculations of the isotope shifts in highly charged ions 
\cite{Tupitsyn_2003, Kozhedub_2010, Gaigalas_2011, Li_2012, Zubova_2014, Zubova_2016}. Here, with the methods 
developed for highly charged ions, we calculate the isotope shifts of the X-ray lines in neutral atoms. 
As is known, the isotope shifts of the energy levels are mainly determined by the finite nuclear size (field shift) and nuclear recoil (mass shift).

The field shift is caused by the difference in the nuclear charge distribution of the isotopes.
The main contribution to the field shift can be calculated in the framework
of the Dirac-Coulomb-Breit Hamiltonian.  The nuclear charge distribution
is usually approximated by the spherically-symmetric Fermi model:
\begin{equation}
\label{rho}
\rho(r,R)=\frac{N}{1+{\rm{exp}}[(r-c)/a]},
\end{equation}
where the parameter $a$ is generally fixed to be $a=2.3/(4{\rm ln}3)$ fm and the parameters $N$ and $c$ are determined using the given value of the root-mean-square nuclear charge radius 
$R=\langle r^2 \rangle^{1/2}$ and the normalization condition: 
$\int{d\vec{r} \rho({r},R)}=1$. 
The potential induced by $\rho(r,R)$ is defined as \\

\begin{equation}
\label{Vn}
 V_{N}(r,R)= - \,4 \pi \, Z \int\limits_{0}^{\infty} {dr' r'^2 \rho (r',R) \frac{1}{r_{>}}},
\end{equation}
where $r_{>}={\rm{{max}}}(r,r')$. 
This potential is used in the Dirac-Coulomb-Breit Hamiltonian to obtain the
relativistic wave functions. The related isotope shifts are evaluated by the
formula:
\begin{equation}
\delta E_{FS}=\langle \psi \mid \sum_{i} \delta V_N(r_i, R) \mid \psi \rangle,
\end{equation} 
where $\delta V_N(r,R)=V_N(r,R+\delta R)-V_N(r,R)$ and $\delta R$ is the 
difference of the rms radii for the isotopes under consideration.

In Tables~\ref{UISK1} and \ref{UISK2} we present the contributions to the
field shifts for the K$\alpha$-lines in $^{235,238}$U 
and $^{233,238}$U, respectively. The total theoretical values are given
by a sum of the Dirac-Fock, Breit, frequency-dependent Breit,
QED, mass shift and electron-correlation contributions.
Expect for the QED correction, all other terms are evaluated
in the same way as the X-ray line energies. 
The QED corrections are determined employing the approach presented in 
Ref. \cite{Zubova_2014}. Namely, this was done by multiplying the s-state QED correction factor taken from 
Refs. \cite{Milstein_2004,Yerokhin_2011} with the nuclear size effect on the total transition energy.

The obtained theoretical results are compared with the related experimental
data from Ref. \cite{Brockmeier_1965}. We note that the K$\alpha$ lines
were indistinguishable in those experiments and, therefore, the 
K$\alpha_1$ and K$\alpha_2$ transition values taken from Ref.
\cite{Brockmeier_1965} are assumed to be the same. The theoretical uncertainty 
is estimated as a doubled quadratic sum of the an uncertainty due to unknown nuclear polarization and deformation effects and a half of the QED contribution. In accordance with the results of Ref. \cite{Zubova_2014}, we have assumed that the uncertainty caused by uncalculated nuclear polarization and deformation effects should be on the level of 1 \% of the corresponding field shift contribution.

Table \ref{UISL} displays the results of the calculations of the L-line isotope shifts,
which are carried out for uranium isotopes with ${A}=238,235$. 
The isotope shifts of these lines are generally
determined in the same way as for the K$\alpha$ lines. The only difference is the neglecting 
the QED contributions for the L${\alpha_2}$ and L${\beta_1}$ lines. As one
can see, there exists a rather large discrepancy between theory and experiment \cite{Makarov_1996} 
for the L${\beta_1}$ line. The reason of this discrepancy is unclear to us. 

In Table \ref{HgISK_202} the individual contributions to the total isotope shifts 
for the K$\alpha$ lines in $^{204,202}$Hg are presented. It can be seen that the total theoretical results 
are in good agreement with the experimental ones \cite{Lee_1978}. The total values of the 
isotope shifts for different pairs of mercury isotopes are selected 
in Table \ref{HgISK201}. The main theoretical uncertainty comes from the nuclear
polarization contribution. It is worth noting that for all isotopes of mercury the theoretical predictions agree with the experimental ones \cite{Lee_1978}.
%%%%%%%%%%%%%%%%%%%%%%%%%%%%%%%%%%%%%%%%%%%%%%%%%%%%%%%%%%%%%%%%%%%%%
\begin{table}
\small
\vspace{1mm}
\caption{\small
Individual contributions to the isotope shift for the K$\alpha$ lines
in $^{235\,,238}$U (in meV) with given values of nuclear charge radii
($^{235}R~=~5.8287$~fm, $^{238}R~=~5.8569~\text{fm}$).}
\vspace{2mm}
\begin{tabular}{c|c|c}
\hline
\hline
\hspace{42mm}
               & \hspace{9mm} K${\alpha_1}$ \hspace{9mm}
               & \hspace{9mm} K${\alpha_2}$ \hspace{9mm} \\[1mm]
\hline
Dirac-Fock                &~1346.35  &~1323.88  \\[-1mm]
Breit                     &~-12.34    &~-12.06\\[-1mm]
Frequency-dependent Breit &~0.07   &~0.12\\[-1mm]
QED                       &~-13.89  &~-13.89 \\[-1mm]
%                          &~(-7.49)  &~(-7.50)\\[-1mm]
Electron correlations + Auger shift  &~-0.17 &~-0.18\\[-1mm]
Mass shift    &-1.70 & -1.39\\
Total theory  &  1318(30) & 1296(30)\\[-0mm]
\hline
\hline
\end{tabular}
\label{UISK1}
\end{table}
%%%%%%%%%%%%%%%%%%%%%%%%%%%%%%%%%%%%%%%%%%%%%%%%%%%%%%%%%%%%%%%%%%%%%%%
\begin{table}
\small
\vspace{1mm}
\caption{\small 
Individual contributions to the isotope shift for the K$\alpha$ lines
 in $^{233 \,,238}$U  (in meV) with given values of nuclear charge radii
 ($^{233}R~=~5.8138~\text{fm}$, $^{238}R~=~5.8569~\text{fm}$).}
\vspace{2mm}
\begin{tabular}{c|c|c}
\hline
\hline
\hspace{42mm}
               & \hspace{9mm} K${\alpha_1}$ \hspace{9mm}
               & \hspace{9mm} K${\alpha_2}$ \hspace{9mm} \\[1mm]
\hline
Dirac-Fock                &   2056.57  &  2022.24~   \\[-1mm]
Breit                     &   -18.86   &  -18.42     \\[-1mm]
Frequency-dependent Breit &   ~0.11 &    ~0.19      \\[-1mm]
QED                       &  -21.20   &  -21.21       \\[-1mm]
%                          &  (-11.43) &  (-11.44)     \\[-1mm]
Electron correlations + Auger shift  &   -0.25   &   -0.26 \\[-1mm]
Mass shift   & -2.86 & -2.34\\[-1mm]
Total theory  & 2014(46)  & 1980(45)\\[0mm]
Experiment \cite{Brockmeier_1965}  &   1800(200) & 1800(200)  \\
\hline
\hline
\end{tabular}
\label{UISK2}
\end{table}
%%%%%%%%%%%%%%%%%%%%%%%%%%%%%%%%%%%%%%%%%%%%%%%%%%%%%%%%%%%%%%%%%%%%%
\begin{table}[ht]
\begin{center}
\small
\vspace{1mm}
\caption{Individual contributions to the isotope shift for the
$L$lines in $^{235,238}$U (in meV) with given values of nuclear charge
radii ($^{235}R~=~5.8287~\text{fm}$, $^{238}R~=~5.8569~\text{fm}$).}
\vspace{2mm}
\begin{tabular}{c|c|c|c|c}
\hline \hline
\hspace{10mm}
               & \hspace{3mm} {\small L${\alpha_2}$} \hspace{3mm}
               & \hspace{3mm} {\small L${\beta_1}$} \hspace{3mm}
               & \hspace{3mm} {\small L${\beta_3}$} \hspace{3mm}
               & \hspace{3mm} {\small L${\beta_4}$} \hspace{3mm} \\[1mm]
\hline
Dirac-Fock   &   -5.608   &   16.863 &   228.750 &   222.565 \\[-1mm]
Breit        &   0.084    &  -0.200   &  -1.444  &  -1.372  \\[-1mm]
Frequency-dependent Breit &  ~0.046   &  -0.003  &  -0.041  &  -0.025 \\[-1mm]
QED                       &  -    &  -   &  -2.203   & -2.194\\[-1mm]
%                          &  (-0.074) &  (-0.065)  &  (-1.551) &  (-1.551) \\[-1mm]
Electron correlations + Auger shift &-0.002 &-0.004 & 0.013  &   0.012 \\[-1mm]
Mass shift & -0.079 & -0.229 & -0.454 &-0.394\\[-1mm]
\hline
Total theory &-5.56(11) & 16.43(35) & 225(5)  &  219(5)  \\
Experiment \cite{Makarov_1996} &-6(2)    & 30(2)    & 253(8)  &  241(10)  \\
\hline
\hline
\end{tabular}
\label{UISL}
\end{center}
\end{table}
%%%%%%%%%%%%%%%%%%%%%%%%%%%%%%%%%%%%%%%%%%%%%%%%%%%%%%%%%%%%%%%%%%%%%
\section{Conclusion}
In this paper we have evaluated the energies and the isotope shifts of the X-ray lines
in neutral atoms using configuration-interaction method in the Dirac-Fock-Sturm
basis in approximation of the barycenter of valence nonrelativistic configuration.
The obtained results are compared with the previous calculations and experiments. 
The comparison demonstrates good agreement of the obtained theoretical results for the 
K lines and the related isotope shifts in uranium and mercury atoms. 
In case of the L lines, 
there exist some discrepancies between theory and experiment for the isotope
shifts in uranium atoms. The discrepancy becomes especially large for the 
L${\beta_1}$ lines. The reason of this discrepancy remains unclear to us.
%%%%%%%%%%%%%%%%%%%%%%%%%%%%%%%%%%%%%%%%%%%%%%%%%%%%%%%%%%%%%%%%%%%%%
%%%%%%%%%%%%%%%%%%%%%%%%%%%%%%%%%%%%%%%%%%%%%%%%%%%%%%%%%%%%%%%%%%
\begin{acknowledgments}
This work was supported by RFBR (Grants No.~18-03-01220, No.~18-32-00275, No.~16-02-00334, No.~17-02-00216) and by SPSU-DFG 
(Grants No. 11.65.41.2017 and No. STO 346/5-1). I.~I.~T. acknowledges financial support from the Ministry of Education and Science of the 
Russian Federation (Grant No. 3.1463.2017/4.6). N.~A.~Z. acknowledges 
financial support from the SPSU (Grant No.~11.42.662.2017), DAAD, and FRRC. We are grateful to the Resource ”Computer Center of SPSU”.  
\end{acknowledgments}
%\nocite{*}
%%%%%%%%%%%%%%%%%%%%%%%%%%%%%%%%%%%%%%%%%%%%%%%%%%%%%%%%%%%%%%%%%%%%%
\begin{table}
\small
\vspace{1mm}
\caption{\small
Individual contributions to the isotope shift for the $K\alpha$ lines
in $^{204 \,,202}$Hg (in meV) with given values of nuclear charge radii
($^{204}R~=~5.4744~\text{fm}$, $^{202}R~=~5.4648~\text{fm}$).}
\vspace{2mm}
\begin{tabular}{c|c|c}
\hline
\hline
\hspace{42mm}
               & \hspace{9mm} K${\alpha_1}$ \hspace{9mm}
               & \hspace{9mm} K${\alpha_2}$ \hspace{9mm} \\[1mm]
\hline
Dirac-Fock                &   -149.116  &  -149.118~   \\[-1mm]
Breit                     &   1.227   &  1.229     \\[-1mm]
Frequency-dependent Breit &   ~-0.007 &    ~-0.007      \\[-1mm]
QED                       &  -1.614   &  -1.619       \\[-1mm]
%                          &  (-1.010) &  (-1.010)     \\[-1mm]
Electron correlations + Auger shift  &  0.098   &   0.102 \\[-1mm]
Mass shift & 1.199 &1.079\\[-1mm]
\hline
Total theory &~-148(3) &~-147(3) \\
Experiment \cite{Lee_1978}&-156(44)&-156(44) \\
\hline
\hline
\end{tabular}
\label{HgISK_202}
\end{table}
%%%%%%%%%%%%%%%%%%%%%%%%%%%%%%%%
\begin{table}
\small
\vspace{1mm}
\caption{\small
Total isotope shifts for the $K\alpha$ lines
 in $^{204\,,202}$Hg,$^{204\,,201}$Hg,$^{204\,,200}$Hg, $^{204,\,199}$Hg, and
$^{204\,,198}$Hg (in meV) with given values of nuclear charge radii taken from Ref. \cite{Angeli_2013}.}
\vspace{2mm}
\begin{tabular}{c|c|c|c}
\hline
\hline
\hspace{42mm}
       & ~~~~~~~~~~~~~~~~~~~~~~~~~~~~ & \hspace{9mm} K${\alpha_1}$ \hspace{9mm}
               & \hspace{9mm} K${\alpha_2}$ \hspace{9mm} \\[1mm]
\hline
$^{204\,, 202}$Hg
& Theory  &~-148(3)~~ &~-147(3)~~\\[-2mm]
& Experiment \cite{Lee_1978} &-156(44)&-156(44) \\[0mm]
$^{204\,,201}$Hg
& Theory &-246(6) &-246(6) \\[-2mm]
& Experiment \cite{Lee_1978} &-286(36)& -286(36) \\[0mm]
$^{204\, ,200}$Hg
& Theory  &-291(7) &-292(7) \\[-2mm]
& Experiment \cite{Lee_1978}&-305(30)& -305(30) \\[0mm]
$^{204\,,199}$Hg
& Theory  &-408(9)&-408(9) \\[-2mm]
& Experiment \cite{Lee_1978} &-425(40)&-425(40) \\[0mm]
$^{204\,,198}$Hg
& Theory  &-424(10)& -424(10)\\[-2mm]
& Experiment \cite{Lee_1978} &-468(44)& -468(44) \\[-1mm]
\hline
\hline
\end{tabular}
\label{HgISK201}
\end{table}
%%%%%%%%%%%%%%%%%%%%%%%%%%%%%%%%
%%%%%%%%%%%%%%%%%%%%%%%%%%%%%%%%%%%%%%%%%%%%%%%%%%%%%%%%%%%%%%%%%%%%%
\clearpage
\section* {\large Appendix A: Dirac-Fock method with the approximation of the
average of relativistic configuration (jj-average) and the average of
nonrelativistic configuration (LS-average)}
\label{sec:LS-average}
\setcounter{equation}{0}
\renewcommand{\theequation}{A\arabic{equation}}
%%%%%%%%%%%%%%%%%%%%%%%%%%%%%%%%%%%%%%%%%%%%%%%%%%%%%%%%%%%%%%%%%%%%%
Let indices $a$ and $b$ enumerate relativistic shells, $A$ and $B$ denote
nonrelativistic shells, $q_a$ and $q_b$ are the numbers of electrons
(occupation numbers) in the shells $a$ and $b$, and $q_A$ and $q_B$ are the numbers
of electrons in the nonrelativistic shells $A$ and $B$, respectively. Thus $A=(n_A l_A)$,
$a=(n_A l_A j_a)= (A \, j_a)$,
and 
$$
q_A = \sum_{a \in A} q_a, \qquad  q_B = \sum_{b \in B} q_b.
$$
First we consider the relativistic average configuration (jj-average).
In this approximation the energy is expressed as \cite{Grant_1970}
\begin{equation}
\begin{array}{lll}
E^{\mathrm{DF}}_{\mathrm{rav}} &=& \displaystyle \sum_{a} q_a \, I_a+
\frac{1}{2}\sum_{a} q_a \, (q_a-1) \, F^{0}(a,a)+
\sum_{a<b} q_a \, q_b \, F^{0}(a,b)
\\[4mm] &+& \displaystyle
\sum_{a}\sum_{k>0} q_a (q_a-1) \, f^k_{aa} \, F^k(a,a) +
\sum_{a<b}\sum_{k} q_a \, q_b \, g^k_{ab} \, G^k(a,b),
\end{array}
\label{E_av}
\end{equation}
where $q_a$ and $q_b$ are the numbers of electrons in the shells $a$ and $b$,
$I_a$ is the one-electron radial integral \cite{Grant_1970}, and $F^{k}(a,b)$
and $G^{k}(a,b)$ are the standard Coulomb and exchange radial integrals
\cite{Grant_1970}, respectively.
The coefficients $ f^k_{a,a}$ and $g^k_{a,b}$ are given by
\begin{equation}
\begin{array}{lll}
f^k_{a,a} &=& \displaystyle
-\, \frac{1}{2} \, \frac{2j_a+1}{2j_a} \,
\frac{\left( C^{k0}_{j_a -\frac{1}{2}, j_a \frac{1}{2}} \right )^2}{2k+1} \,=\,
- \, \frac{1}{4} \,\frac{2j_a+1}{2j_a} \, \Gamma^k_{j_a,j_a},
\\[5mm] \displaystyle
g^k_{a,b} &=&  \displaystyle
- \, \frac{\left( C^{k0}_{j_a -\frac{1}{2}, j_b \frac{1}{2}} \right )^2}{2k+1} \,=\,
\,-\, \frac{1}{2}  \, \Gamma^k_{j_a,j_b}.
\end{array}
\label{AB_coef2}
\end{equation}
Where $\Gamma^k_{j_a,j_b}$ are the coefficients introduced in Ref. \cite{Grant_1970},
\begin{equation}
\Gamma^k_{j_a,j_b} = 
2 \, \left (
\begin{array}{llll}
 j_a & j_b & k \\
 \frac{1}{2} & -\frac{1}{2}  & 0
\end{array} \right )^2 \,.
\end{equation}
The procedure of the relativistic configurational average is meaningful only
when the jj-coupling dominates, that obviously is not true for most of 
neutral atoms. Furthermore, the use of the pure jj-coupling scheme leads to a 
wrong nonrelativistic limit. For this reason it is reasonable to use the averaging
over all the jj-configurations arising from a valence nonrelativistic
configuration in the calculations of neutral atoms. Starting with equation
(\ref{E_nav1}) we obtain the following energy expression in the nonrelativistic
configurational average (LS-average)
\begin{equation}
\begin{array}{lll}
E^{\mathrm{DF}}_{\mathrm{nav}} &=& \displaystyle \sum_{a}{\widetilde q_a \, I_a}+
\frac{1}{2}\sum_{a}{\widetilde q_a (\widetilde q_a -w_A)F^{0}(a,a)}+
\sum_{a<b}{\widetilde q_a  \widetilde q_b \, \omega_{AB} \, F^{0}(a,b)}
\\[4mm] &+& \displaystyle
\sum_{a}\sum_{k>0}{\widetilde q_a (\widetilde q_a -w_A) \, f^k_{aa}F^k(a,a)}
+\sum_{a<b}\sum_{k} \widetilde q_a \, \widetilde q_b \, \omega_{AB} \,
g^k_{ab} \, G^k(a,b),
\end{array}
\label{E_nav2}
\end{equation}
where the parameters $\widetilde q_a$, $w_a$, and $\omega_{A B}$ are defined as 
\begin{equation}
\widetilde q_a=\frac{2j_a+1}{4l_A+2} \, q_{A} \,, \qquad
w_a=\frac{q_{A}-\widetilde{q_a}+2j_a}{4l_A+1},\\
\end{equation}
\begin{equation*}
\omega_{AB}=
\begin{cases} \displaystyle
\frac{4L_a+2}{4L_a+1} \, \frac{q_A-1}{q_A}, \qquad A = B.
\\  \displaystyle
\qquad 1 \qquad \qquad \qquad A \neq B.
\end{cases}
\end{equation*}
Here $q_{A}$ is the total number of electrons in the nonrelativistic
shell $A=n_a l_a$.

The expression (\ref{E_nav2}) can be rewritten in the same form as the
nonrelativistic expression for the energy in the Hartree-Fock method
\cite{Fischer_1977},
\begin{equation}
\begin{array}{lll}
E^{\rm DF}_{\rm nav} &=&  \displaystyle
\sum_{A} q_{A} \, \overline I_{A} +
\frac{1}{2} \, \sum_{A} \, q_{A} \,(q_{A}-1) \, \overline F^{~0}(A,A) +
\sum_{A < B} \, q_{A} \, q_{B} \, \overline F^{~0}(A,B)
\\[4mm] &+& \displaystyle
\sum_{A} \, \sum_{k>0} \,q_{A} \,(q_{A}-1) \, \overline f^k_{A,A} \,
\overline F^k(A,A) +
\sum_{A < B} \sum_{k} q_{A} \, q_{B} \, \, \overline g^k_{A,B} \,,
\overline G^k(A,B) \,,
\end{array}
\label{E_nav3}
\end{equation}
where  $\overline F^k(A,B)$ and $\overline G^k(A,B)$ are effective mean values
of the radial integrals defined as\\
\begin{equation}
\overline F^{~0}(A,B) = \left \{
\begin{array}{ll} \displaystyle
\sum_{j_a \in A} \sum_{j_b \in B}
\frac{(2j_a+1-\delta_{j_a,j_b})(2j_b+1)}{(4l_A+2)(4l_A+1)} \, F^0(a,b) \qquad & A=B
\\[5mm] \displaystyle
\sum_{j_a \in A} \sum_{j_b \in B}
\frac{(2j_a+1)(2j_b+1)}{(4L_a+2)(4L_b+2)} \, F^0(a,b) \quad & A \ne B \, 
\end{array} \right .
\end{equation}
for $k = 0$ and
\small
\begin{equation}
\overline G^{~k}(A, B)  = \frac{1}{2} \sum_{j_a \in A} \sum_{j_b \in B}
(2j_a+1) (2j_b+1)
\left \{
\begin{array}{llll}
j_a & j_b & k \\
l_b & l_a & \frac{1}{2}
\end{array} \right \}^2 
G^k(a, b) \,,
\quad
\overline F^{~k}(A, A) = \overline G^{~k}(A, A) \,
\label{int_G1}
\end{equation}
\normalsize
for $k > 0$.

In the nonrelativistic limit, the integrals   $\overline F^k(A,B)$ and
$\overline G^k(A,B)$ tend to the corresponding nonrelativistic radial
integrals defined in the nonrelativistic Hartree-Fock method \cite{Fischer_1977}.

The coefficients $\overline f^k_{A,A}$ and $\overline g^k_{a,b}$ coincide
with the corresponding coefficients defined  in the nonrelativistic Hartree-Fock method
in the approximation of the center of gravity,
\small
\begin{equation}
\begin{array}{lll}
\overline f^k_{A,A} &=& \displaystyle -\, \frac{1}{4} \, \frac{4l_A+2}{4l_A+1} \, 
\frac{\left( C^{k0}_{l_A 0, l_A 0} \right )^2}{2k+1} \,,
\\[4mm] \displaystyle
\overline g^k_{a,b} &=&  \displaystyle
\,-\, \frac{1}{2} \, \frac{1}{2k+1} \, \left( C^{k0}_{l_A 0, l_B 0} \right )^2 \,.
\end{array}
\label{AB_coef1}
\end{equation}
\normalsize
%%%%%%%%%%%%%%%%%%%%%%%%%%%%%%%%%%%%%%%%%%%%%%%%%%%%%%%%%%%%%%%%%%%%%

%%%%%%%%%%%%%%%%%%%%%%%%%%%%%%%%%%%%%%%%%%%%%%%%%%%%%%%%%%%%%%%%%%%%%
\end{document}